\documentclass[aps,prm,twocolumn,superscriptaddress]{revtex4-1}
\usepackage{graphicx,bm,amssymb,amsmath,xcolor,pifont,soul}
\usepackage[linktocpage=true,colorlinks=true,pdfborder={0 0 0},linkcolor=blue,citecolor=red,filecolor=yellow,urlcolor=blue,bookmarks,pdfauthor={},]{hyperref}
\usepackage{ulem}
\usepackage{multirow}
\newcommand{\HP}[1]{\textcolor{black}{#1}}

\begin{document}

\title{Intrinsic Berry phase contribution to Hall conductivity in CoS$_2$}
\author{Tamal Kumar Dalui}
\affiliation{DOD Center of excellence for Advanced Electro-Photonics with 2D Materials, Morgan State University, Baltimore, Maryland-21251, USA}
\author{Hari Paudyal}
\affiliation{Department of Physics and Astronomy, University of Iowa, Iowa City, Iowa 52242, USA}
\author{Durga Paudyal}
\affiliation{Department of Physics and Astronomy, University of Iowa, Iowa City, Iowa 52242, USA} 
\author{Ramesh C Budhani}
\email{ramesh.budhani@morgan.edu}
\affiliation{DOD Center of excellence for Advanced Electro-Photonics with 2D Materials, Morgan State University, Baltimore, Maryland-21251, USA}

\date{\today}

\begin{abstract}
In Weyl semi-metals, the conduction and valence bands intersect at distinct points on the Brillouin zone (Weyl points), which act as monopoles of Berry curvature in momentum space. This nontrivial band topology, identified from electronic structure calculations, gives rise to various exotic magneto-transport properties. 
Hybrid functional calculations that incorporate a portion of exact exchange, magneto-transport measurements, and  temperature-dependent resistivity confirm nontrivial band topology and half-metallicity in CoS$_2$ of magnetic ordering temperature $T_{\rm C} \approx 120~\mathrm{K}$. \HP{However, electronic structure calculations also show that application  of small strain transforms this half metallic character to the metallic}. Interestingly, the magnetoresistance (MR) of the CoS$_2$ films is characterized by a reentrant weak localization above a critical field at $T \leq 60\,\mathrm{K}$ and a negative to positive transition in MR as the $T$ goes from $<T_{\rm C}$ to $>T_{\rm C}$. Experimental observation of anomalous Hall resistivity and \textit{ab initio} computed band structure, Berry curvature, and Hall conductivity ($\sigma_{xy}$) demonstrate that the $\sigma_{xy}$ in CoS$_2$ is primarily driven by the intrinsic Karplus–Luttinger contribution, often linked to Berry phase physics. 
\end{abstract}
\maketitle

\section{Introduction}
Recent studies have revealed that CoS$_2$, which crystallizes in the pyrite structure$-$a material long studied for its itinerant ferromagnetism and potential for half-metallicity$-$hosts Weyl fermions and Fermi arc surface states close to the Fermi level (E$_{\rm F}$), as well as topological nodal lines below the E$_{\rm F}$~\cite{schroter2020weyl}. In Weyl semimetals (WSMs), the conduction and valence bands touch at isolated points known as Weyl points, which act as monopoles of Berry curvature in momentum space. This nontrivial band topology results in various exotic transport properties, such as a giant anomalous Hall effect (AHE), negative magnetoresistance, planar Hall effect, and anomalous Nernst effect. These features in the magneto-transport are key indicators of the Weyl character of electronic bands~\cite{li2020giant, li2017negative, gopal2020observation, reichlova2018large}.

Further, CoS$_2$ is an interesting example of the recently discovered class of experimentally verified magnetic topological metals (MTM)~\cite{liu2019magnetic, belopolski2019discovery, morali2019fermi}, which may potentially offer new mechanisms of spin-to-charge conversion in magnetic heterostructures~\cite{zhang2019spin, brown2005magnetization}. The discovery of the magnetic topological metals is of broader interest for fundamental science, providing a platform to realize axion insulators~\cite{wan2011topological}, the intrinsic AHE~\cite{burkov2014anomalous}, and the anomalous fractional quantum Hall effect~\cite{wang2020fractional}. Recent research has also shown a strong connection between AHE and the Berry curvature of electronic states, particularly in topological semimetals (TSMs) and magnetic Weyl semimetals, where Weyl nodes near \HP{the E$_{\text{F}}$} contribute to large intrinsic AHE~\cite{burkov2014anomalous}. Single-crystalline CoS$_2$ shows a large anomalous Hall conductivity (AHC), which has been attributed to its half-metallic nature and the breaking of time-reversal symmetry~\cite{choi2024tunable}. The large positive magnetoresistance seen in CoS$_2$ is indicative of Weyl nodes near E$_{\rm F}$, which acts as a source of Berry curvature~\cite{zhang2022scaling}. Additionally, CoS$_2$ exhibits transport properties influenced by spin, as evidenced by magnetoresistance and spin polarization~\cite{jarrett1968evidence, adachi1981hall, wang2004spin, wang2005c}.  Polycrystalline CoS$_2$ also shows technologically important properties such as high thermoelectric power from room temperature to high temperature regimes~\cite{hebert2013transport}, and high electrocatalytic activity suitable for the enhanced performance of quantum dot-sensitized solar cells ~\cite{faber2013earth}.

In this study, we investigate the AHE of highly textured thin films of CoS$_2$. A ferromagnetic state develops in these samples at a critical temperature of 120 K, which is consistent with earlier measurements of the Curie temperature in both bulk polycrystalline samples and nanoparticles of CoS$_2$~\cite{wang2004spin, kumar2014synthesis}. A quadratic scaling relationship between the anomalous Hall resistivity ($\rho^{AHE}_{xy}$) and longitudinal resistivity ($\rho_{xx}$) is observed, which indicates that the primary contribution to the large AHE in these samples comes from the band structure driven intrinsic mechanism. The electronic band structure, Berry curvature, and Hall conductivity calculations support the intrinsic Berry phase contribution to Hall conductivity and half metallic character of CoS$_2$.

\section{Experimental Details}
The precursor for the growth of CoS$_2$ is a 10 nm thick film of cobalt deposited by dc magnetron sputtering at the rate of 0.1~\AA/s on 2 – inch diameter thermally oxidized silicon wafers with 300 nm thick SiO$_2$. The Co films were deposited at 5 mTorr ultrahigh purity Ar gas pressure in a vacuum system of base pressure $\sim$ 1.5 x 10$^{-7}$ torr. The thermal sulfurization of the Co film in the presence of sulfur vapors was carried out using an atmospheric pressure chemical vapor deposition reactor which consists of a two-zone tubular furnace. Before initiating the thermal ramp cycle, the reactor tube was evacuated with a dry pump and flushed several times with an Ar/H$_{\rm 2}$ mixture. A constant flow rate of 25 standard cubic centimeter per minute of Ar (with 5 (\%) H$_{\rm 2}$) was maintained during the reaction cycle. The temperature of the center zone (Zone 1), containing the Co film, was raised to 500 $^{\circ}$C, while the temperature of the sulfur-containing zone (Zone 2) was ramped up to 250 $^{\circ}$C. These temperatures were held constant for 2 hours for the sulfurization reaction to complete and then, the furnace was rapidly cooled to room temperature by opening its top lid. The crystallographic structure of the film was measured with X-ray diffraction (XRD) using a Rigaku smart lab X-ray diffractometer with Cu-$K\alpha$1 radiation. The Hall resistivity and magnetoresistance of the CoS$_2$ films were measured down to 10 K in a physical property measurement system equipped with a 9-tesla superconducting magnet.

\begin{figure}[t]
\centering
\includegraphics[width=0.49\textwidth]{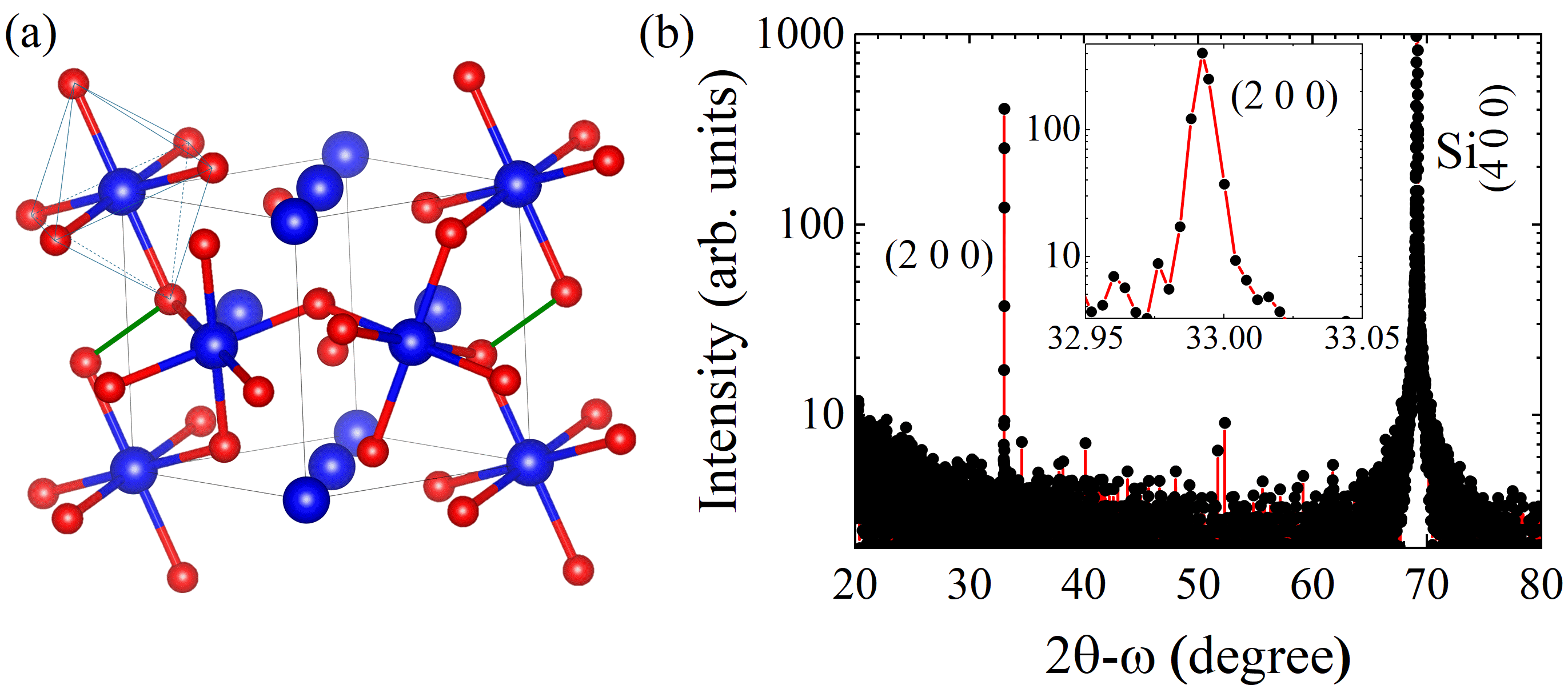}\hfill
\caption {(a) Crystal structure of CoS$_2$ showing octahedrally coordinated Co atoms (Blue balls). (b) X-ray diffraction ($2\theta-\omega$) scan of the 10~nm thick CoS$_2$ film collected at ambient temperature. Inset shows the magnified view of the scan centered at ($200$) reflection.} 
\label{fig-crystal}
\end{figure}

\section{Computational Details}
First-principles electronic structure calculations are performed using the Vienna \textit{Ab initio} Simulation Package~\cite{furthmuller1996dimer, kresse1996efficiency} using projector augmented wave
pseudopotentials~\cite{blochl1994projector} and Perdew-Burke-Ernzerhof (PBE) functionals~\cite{perdew1996o} within the generalized gradient approximation. A plane wave energy cutoff of 500~eV, a smearing value of 0.1~eV, and a $\Gamma$-centered 12 $\times$  12 $\times$ 12 Monkhorst-Pack \textbf{k}-mesh are used in structural relaxation and electronic structure calculations. The unit-cell structure contains four formula units, and all calculations, unless otherwise specified, are performed for this structure. The self-consistent energy convergence criterion of 10$^{-6}$~eV is used in the calculations of atomic and electronic structures, while the atomic positions and lattice constants are relaxed until the maximum force on each atom was less than 10$^{-4}$~eV/\AA. Spin-orbit coupling (SOC) effects are incorporated into the self-consistent field calculations, with the magnetization oriented along the $c$-axis. The underestimated band gaps obtained with PBE are corrected using hybrid functional calculations, which partly incorporate the exchange-correlation potential of standard density functional theory with the Hartree-Fock exchange~\cite{heyd2003hybrid}. Maximally localized Wannier functions, including Co $d$ orbitals and S $p$ orbitals, are used to describe the electronic structure and calculate AHC and Berry curvature using Wannier90~\cite{Pizzi2019s} and Wannier tool~\cite{WannierTools2017} packages.

\begin{figure}[t]
\centering
\includegraphics[width=0.49\textwidth]{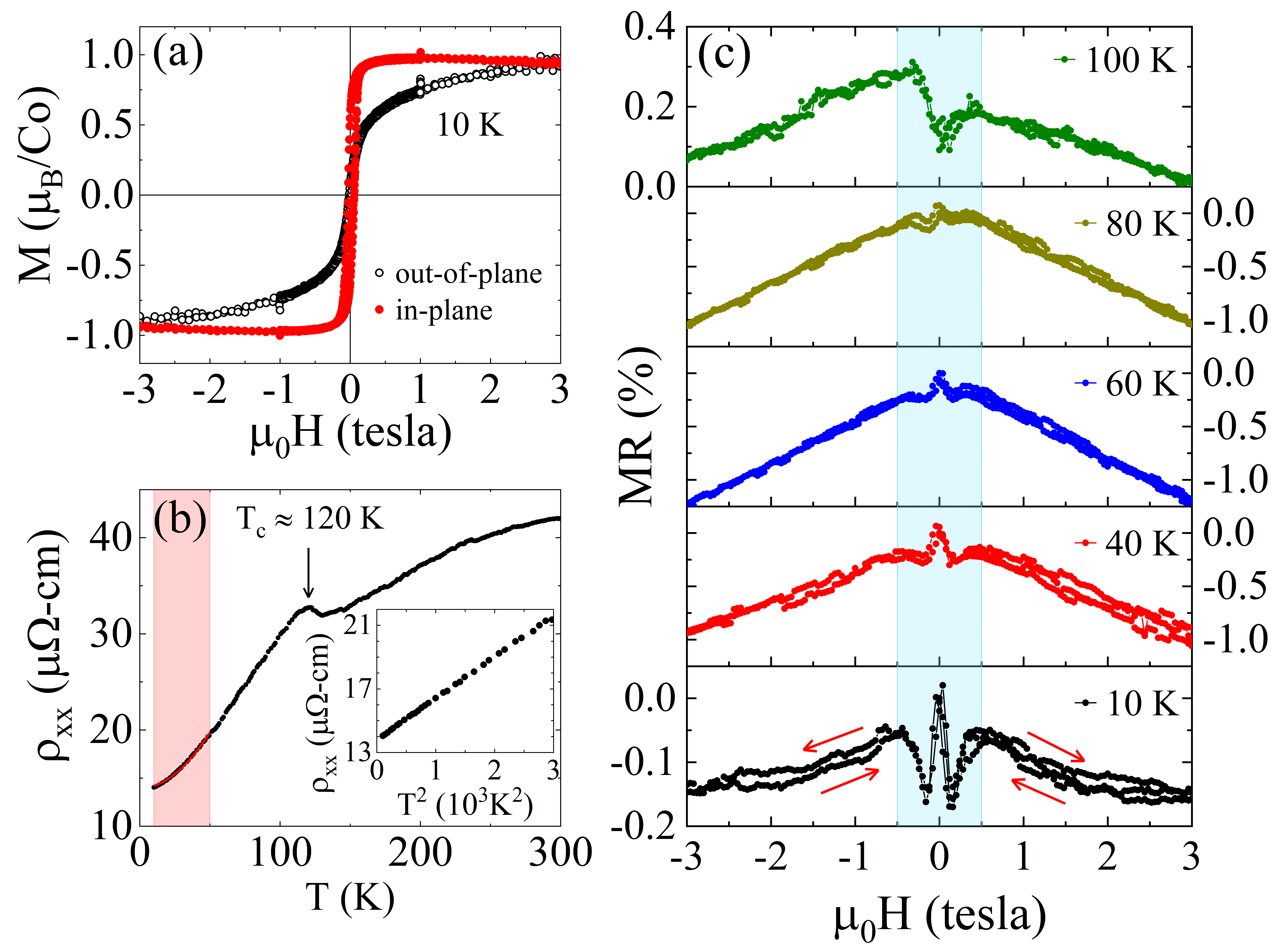}\hfill
\caption {(a) The isothermal magnetization at 10 K, measured with the magnetic field applied out-of-plane and in-plane of the sample. (b) Temperature dependence of the zero-field resistivity of the CoS$_2$ film, inset shows the T$^2$ dependence of the resistivity over the temperature range 10 to 50~K. (c) The MR (\%) vs magnetic field for different temperatures $T<$ T$_C$.} 
\label{fig1}
\end{figure}

\section{Results and Discussion}
\subsection{Crystal Structure, magnetic ordering and magnetoresistance}

Figure \HP{\ref{fig-crystal} shows the crystal structure and} the $\omega-2\theta$ XRD scan of a 10~nm thick CoS$_2$ film. The pattern is characterized by the strong (400) reflection of the silicon substrate at $2\theta$ = 69.2$^{\circ}$ and a sharp but low intensity line centered at $2\theta$ = 32.9$^{\circ}$. Since CoS$_2$ crystallizes in the cubic $Pa\overline{3}$ space group~\cite{hebert2013transport} structure with lattice parameters $a$ = 5.59~\AA, this sharp diffraction line can be indexed as the (200) reflection of the cubic CoS$_2$, which is in agreement with earlier measurements on CoS$_2$ films~\cite{faber2013earth, song2019construction}. \HP{In addition, our theoretically optimized lattice constant of CoS$_2$ after geometrical relaxation is 5.51~\AA, which compares well with the experimental value. The structure of the CoS$_2$ is unique in the sense that the Co atoms (Blue balls) form the vertices of a cube and S atoms are placed in such a way to form S-S dimers, leading to a robust high symmetry structure (Fig.~\ref{fig-crystal}(a))}. 

\begin{figure}[!t]
\centering
\includegraphics[width=0.49\textwidth]{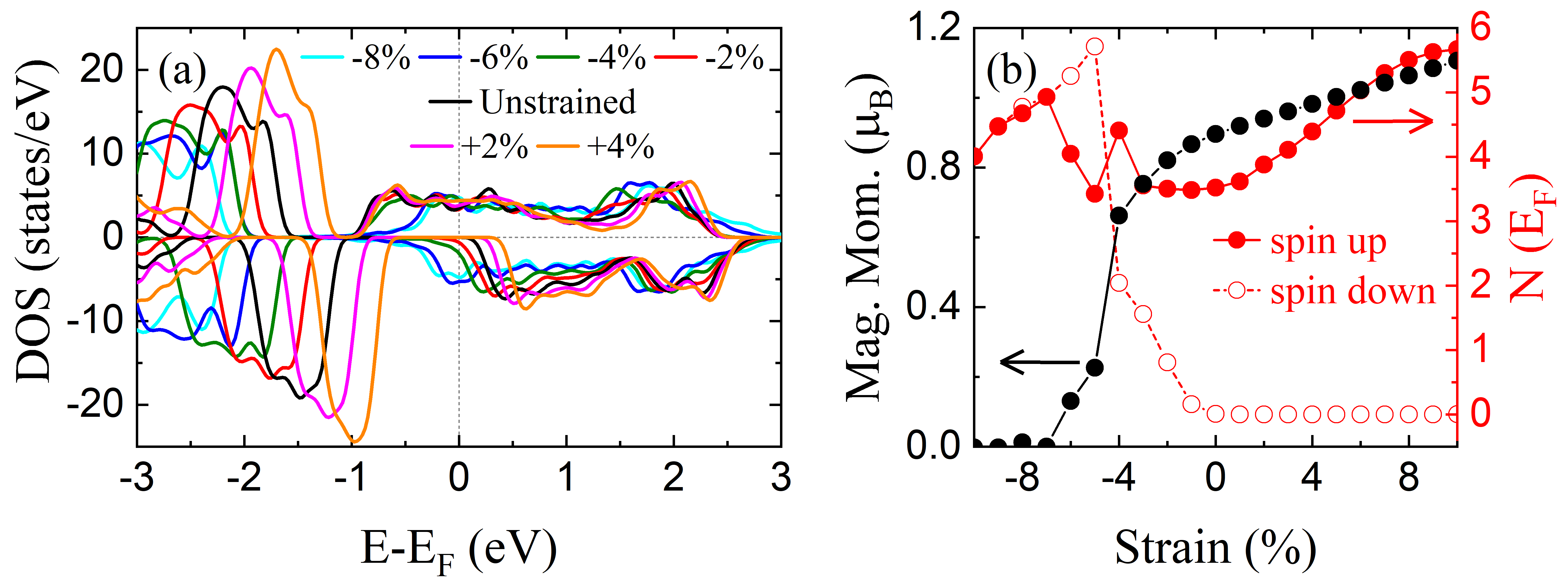}\hfill
\caption {(a) Density of states (DOS) with different compressive and tensile strain. (b) Variation of the magnetic moment (left axis) and integrated DOS up to the Fermi level (right axis) with different compressive and tensile strain.}
\label{fig-dos}
\end{figure}

Figure~\ref{fig1}(a) \HP{shows} the isothermal magnetization data at 10~K, measured with the magnetic field oriented in two distinct direction; parallel to sample normal and parallel to plane. The saturation moments $M_{\rm S}$  are approximately 0.92 ± 0.05 $\mu_{\rm B}$ per Co atom for the out-of-plane geometry and 0.95 ± 0.05 $\mu_{\rm B}$ per Co atom for the in-plane geometry. \HP{These moments compare well with the calculated values of 1.26 $\mu_B$ (1.20 $\mu_B$ spin and 0.06 $\mu_B$ orbital) and those reported in the literature~\cite{schroter2020weyl, wang2004spin}. The magneto-crystalline anisotropy energy (MAE) (E$_{100}$ - E$_{001}$) calculations show slightly different values, as usually observed in transition metal based cubic materials~\cite{coey2010magnetism}. The magnetization data of Fig.~\ref{fig1}(a) reveal that the ferromagnetic properties of CoS$_2$ are anisotropic, with an in-plane anisotropy energy density of $\approx$~11.2~$erg /cm^{3}$.} The zero-field longitudinal resistivity ($\rho_{xx}$) of the films was measured over a temperature range of 10 to 300 K (Fig.~\ref{fig1}(b)). 
The $\rho_{xx}$(T) exhibits a metallic behavior from 300~K down to the lowest temperature, with a hump-like feature at $\approx$~120~K. Below 50 K, the resistivity follows a $T^2$ dependence, as demonstrated in the inset of Fig.~\ref{fig1}(b). This $T^2$ dependence has been previously observed in potential half-metal systems~\cite{akimoto2000observation}. \HP{Our theoretical calculations reveal that CoS$_2$ exhibits a half-metallic nature. However, an application of compressive strain (above 2\%) induces a transition to a fully metallic state (Fig.~\ref{fig-dos}(a)), exhibiting a minority-spin electron pocket at the Brillouin zone corner. Further, the ferromagnetism collapses below 5\% compressive strain with nearly equal integrated spin up and spin down density of states (DOS) up to the Fermi level (Fig.~\ref{fig-dos}(b)). This suggests that the electronic structure is highly sensitive to lattice parameters, with strain modifying the band structure and potentially closing the spin gap. This finding helps to resolve the controversy in previous works by demonstrating that small structural variations can significantly impact electronic properties of CoS$_2$~\cite{schroter2020weyl, yamamoto1999re, shishidou2001effect}.}

\begin{figure}[!t]
\centering
\includegraphics[width=0.45\textwidth]{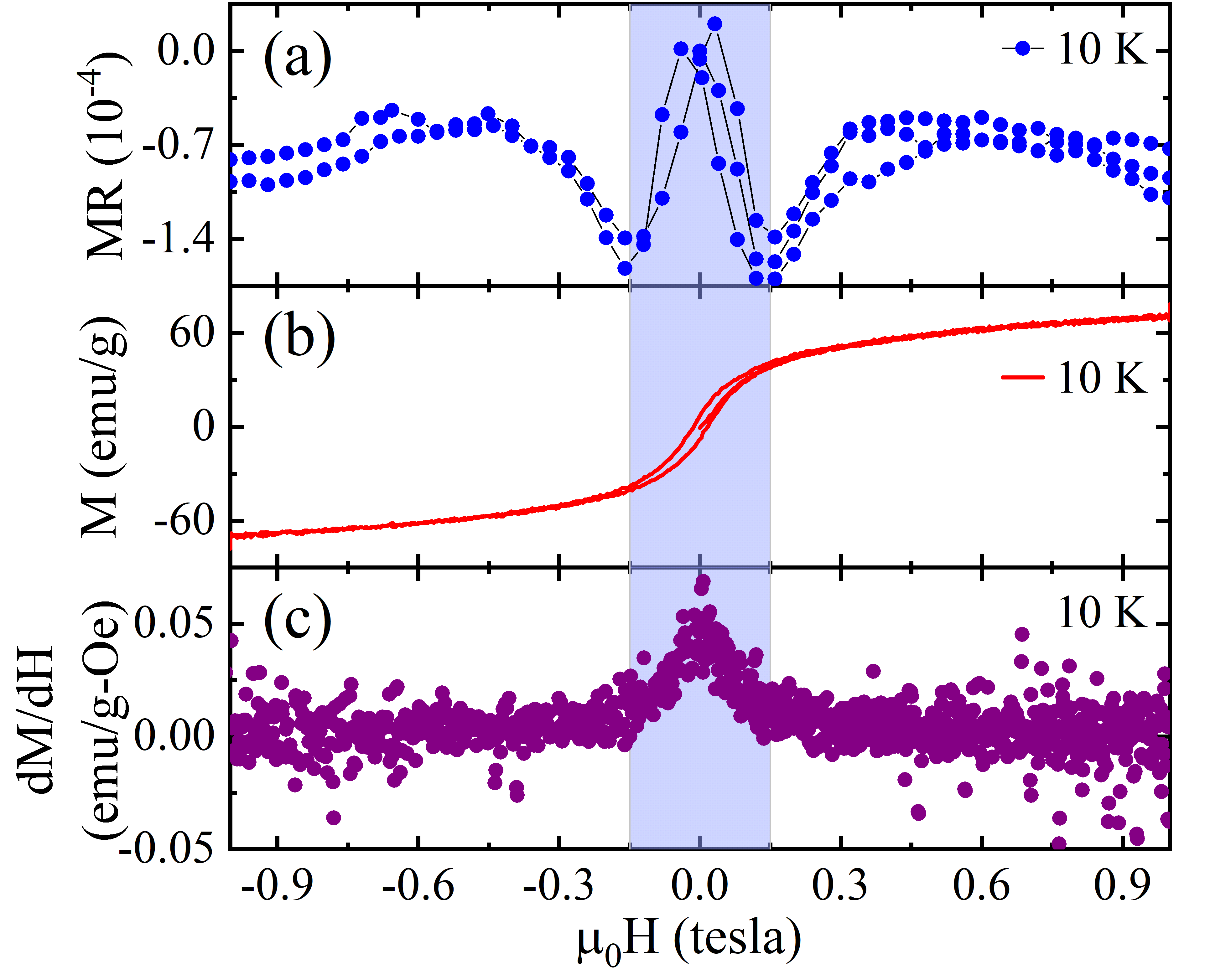}\hfill
\caption {(a) The MR vs magnetic field for 10~K. (b) The static magnetization measured as a function of magnetic field at 10~K. (c) The dM/dH as a function of magnetic field at 10~K.}
\label{fig2}
\end{figure}

The magnetoresistance and Hall resistivity of the film have been measured at several temperatures in the field range of -3 to +3 tesla applied perpendicular to the plane of the sample and the direction of current. The out-of-plane field magnetoresistance is calculated using the relation
\begin{equation}
MR= [\rho_{xx}(H) - \rho_{xx}(0)]/\rho_{xx}(0)
\label{eq1}
\end{equation}
where $\rho_{xx}(H)$ and $\rho_{xx}(0)$ are the resistivity of the sample at field $H$ and in zero-field respectively.
The MR (\%) curves at each temperature below the T$\rm {_C}$  starting from the lowest temperature are shown in Fig.~\ref{fig1}(c). The blue color band in the figure marks the field range $\pm$0.5 tesla, where there is a peak at zero field, followed by a cusp on either side of the peak. The intensity of the zero-field peak decreases as the temperature is raise. This particular feature is suggestive of a magnetic field induced suppression of weak localization in this system of a weak spin orbit coupling ~\cite{lee1985disordered}. Such quantum interference effects are more likely to occur at low temperatures and eventually vanish with the increasing temperature. An intriguing feature of the data shown in Fig.~\ref{fig1}(c) is the cusp in MR in the vicinity of $\mu_{0}H$ $\leq$ 0.15 tesla. While the negative MR (\%) seen at $\mu_{0}H$ $\geq $ 0.5 tesla and $T<$ T$_{\rm C}$ can be interpreted as the field suppression of electron-magnon scattering~\cite{akimoto2000observation, yamada1972negative}, which is present in all conventional ferromagnets, origin of the cusp needs further discussion. For this purpose, Fig.~\ref{fig2}(a) shows an expanded version of the MR data measured at 10 K. The drop in the magnitude of MR (increase in resistivity) with increasing field in the field range of 0.15 to 0.3 tesla is suggestive of an enhance backscattering of charge carriers by a precipitous growth of partially oriented magnetic domains in the system just before the magnetic saturation is reached. To support this argument, Fig.~\ref{fig2}(b) shows the out-of-plane magnetization loop of the film together with the dM/dH in Fig.~\ref{fig2}(c). It is noted that the cusp in MR is located at a point where the magnetization is yet to saturate fully. Clearly, its origin appears to be a result of the competition between magnetic field induced suppression of weak localization, which is dominant at the lower fields and the scattering of charge carriers by randomly oriented domains that causes a positive MR~\cite{akimoto2000observation}.

\begin{figure}[t]
\centering
\includegraphics[width=0.49\textwidth]{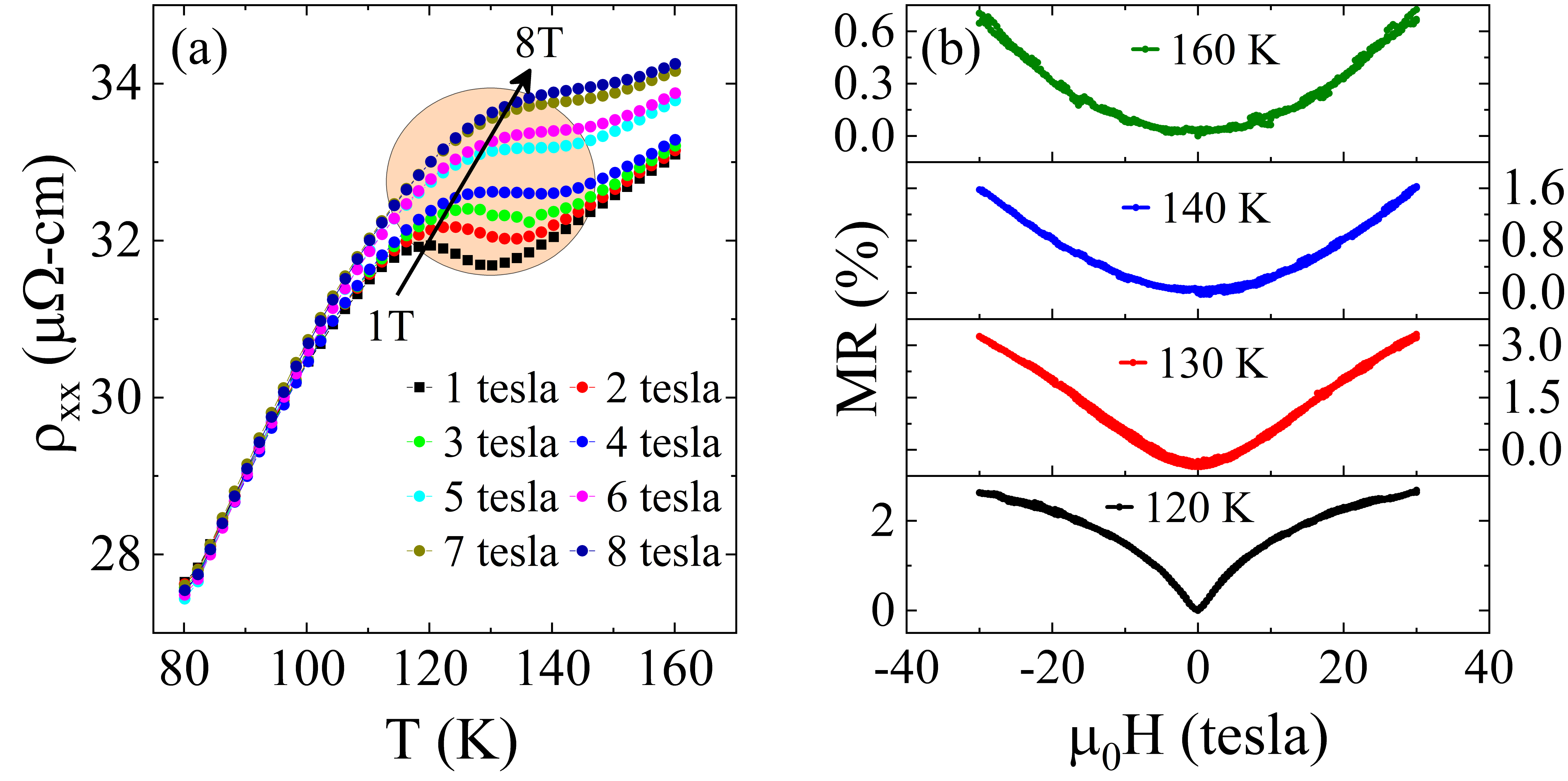}\hfill
\caption {(a) Temperature dependence of the resistivity in the vicinity of T$\rm {_C}$ for different magnetic fields applied perpendicular to the plane of the sample. (b) The MR (\%) vs Magnetic field for different $T>$ T$\rm {_C}$.}
\label{fig4}
\end{figure}

\begin{figure}[t]
\centering
\includegraphics[width=0.49\textwidth]{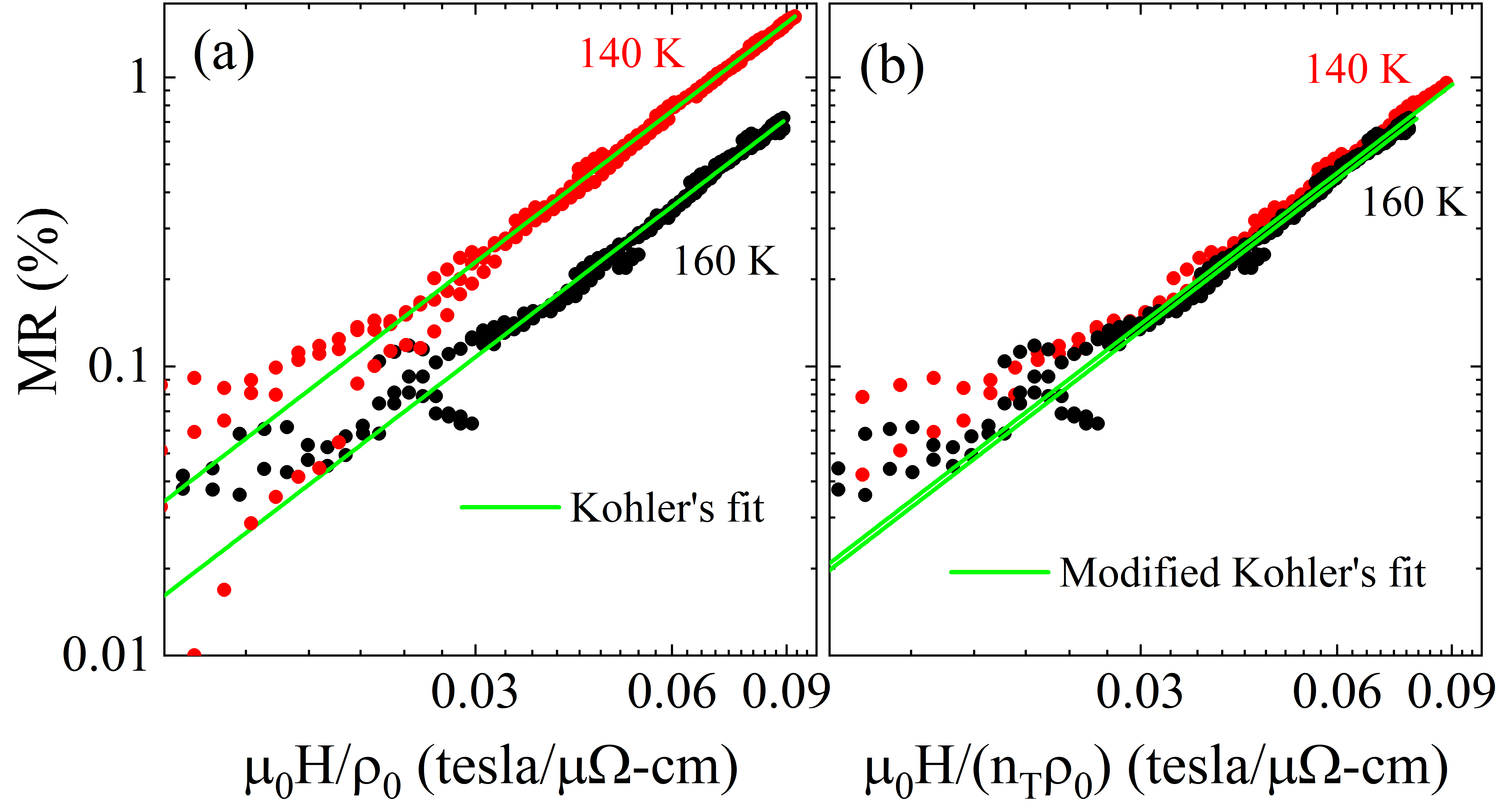}\hfill
\caption {(a) The MR (\%) as a function of $\mu_{0}H/\rho_0$ for two different temperatures, and their corresponding Kohler's fitting is shown by green curves. (b) The MR (\%) as a function of $\mu_{0}H/n_{T}\rho_0$ for two different temperatures, and their corresponding modified Kohler's fitting is shown by green curves.}
\label{fig5}
\end{figure}

The temperature dependence of $\rho_{xx}$ at several fields is presented in Fig.~\ref{fig4}(a), where at $T>$ T$_{\rm C}$  $\approx$~120~K, the curves show a significant upward shift with the increasing field, indicating a large positive magnetoresistance. However, at $T<$ 120 K the dependence of the resistivity on field is miniscule and negative. Further, the field dependence of magnetoresistance at at $T>$ T$_{\rm C}$ shown in Fig.~\ref{fig4}(b) revels that it is posiitive at all fields. The field and temperature dependence of MR in a broad class of materials, including metals, doped oxides, semiconductors, and topological materials, can be understood within the framework of Kohler’s rule~\cite{xu2021extended}. In its simplest form, it states that the MR should follow a scaling law of the form:
 
\begin{equation}
{\rm MR} = \alpha(\mu_{0}H/\rho_{0})^m
\label{eq2}
\end{equation}

where $H$ is the applied magnetic field and $\rho_{0}$ the zero field resistivity at a given temperature. 
The scaling law predicts that all MR curves measured at different temperatures should fall into a single universal curve. Figure ~\ref{fig5}(a) shows the MR data of 140 and 160 K plotted in the framework of Eq.~\ref{eq2}. However, the data deviate from this scaling law. One of the primary reasons for this significant deviation is the omission of temperature-dependent variations in $\rho_{0}$. To address this, the data have been replotted using the modified Kohler’s rule ~\cite{xu2021extended}, which is expressed as 
\begin{equation}
{\rm MR} = \alpha(\mu_{0}H/n_{T}\rho_{0})^m
\label{eq}
\end{equation}
Where $n_{T}$ represents the carrier density renormalized by the carrier effective mass, the values of $n_{T}$ obtained from this scaling are 0.77 at 140 K and 0.81 at 160 K, respectively. Fig.~\ref{fig5}(b) clearly shows a better agreement with Eq.~\ref{eq}. Above 160 K, the MR is approximately zero, so no information on MR can be obtained beyond this temperature.

\subsection{Anomalous Hall effect and Magnetization}

\begin{figure}[ht!]
\centering
\includegraphics[width=0.49\textwidth]{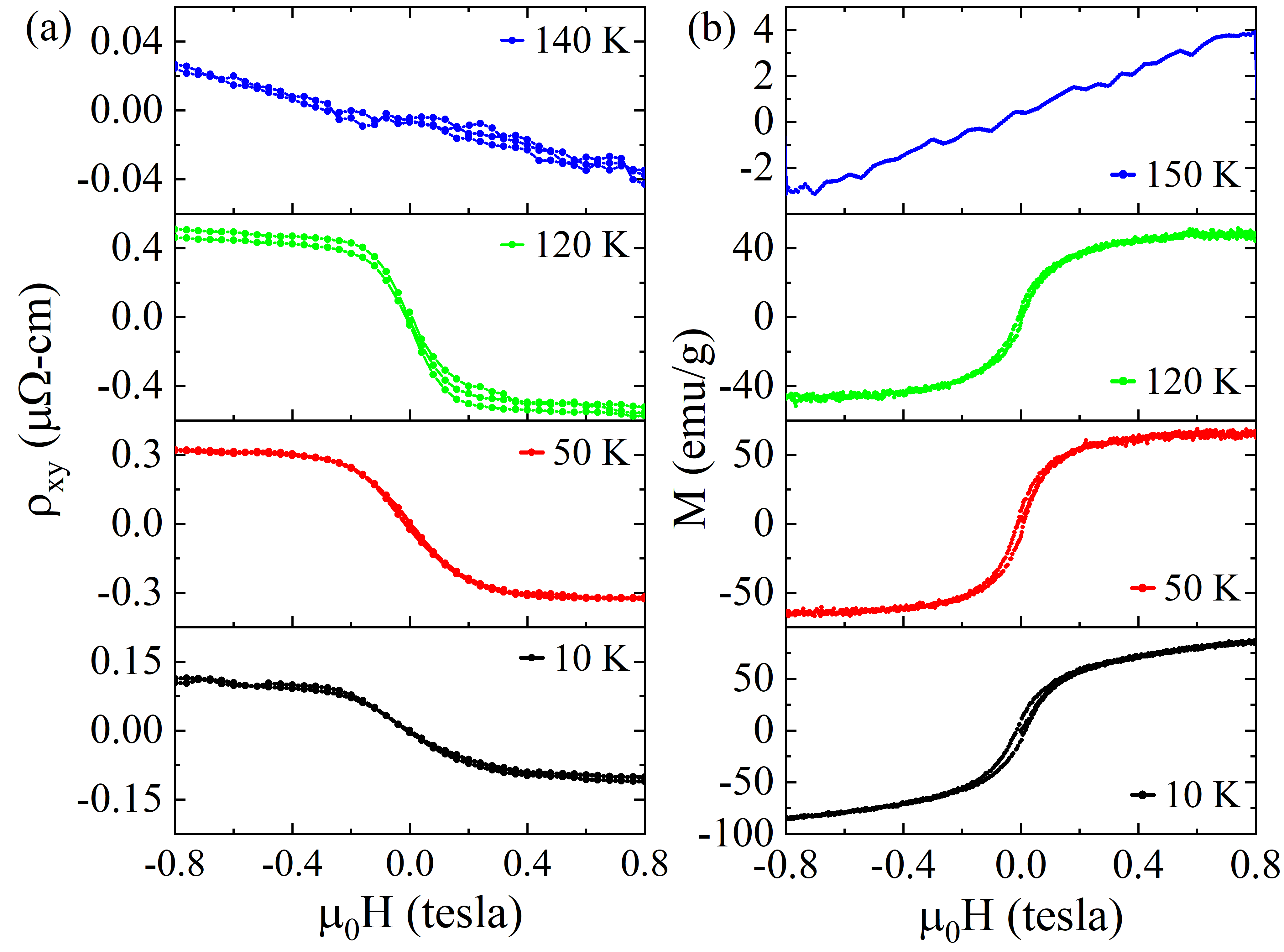}\hfill
\caption {(a) Magnetic field dependence of the Hall resistivity at different temperatures, from 10 to 140~K. (b) The static magnetization measured as a function of magnetic field at the same set of temperatures as in (a).}
\label{fig6}
\end{figure}

\begin{figure}[ht!]
\centering
\includegraphics[width=0.49\textwidth]{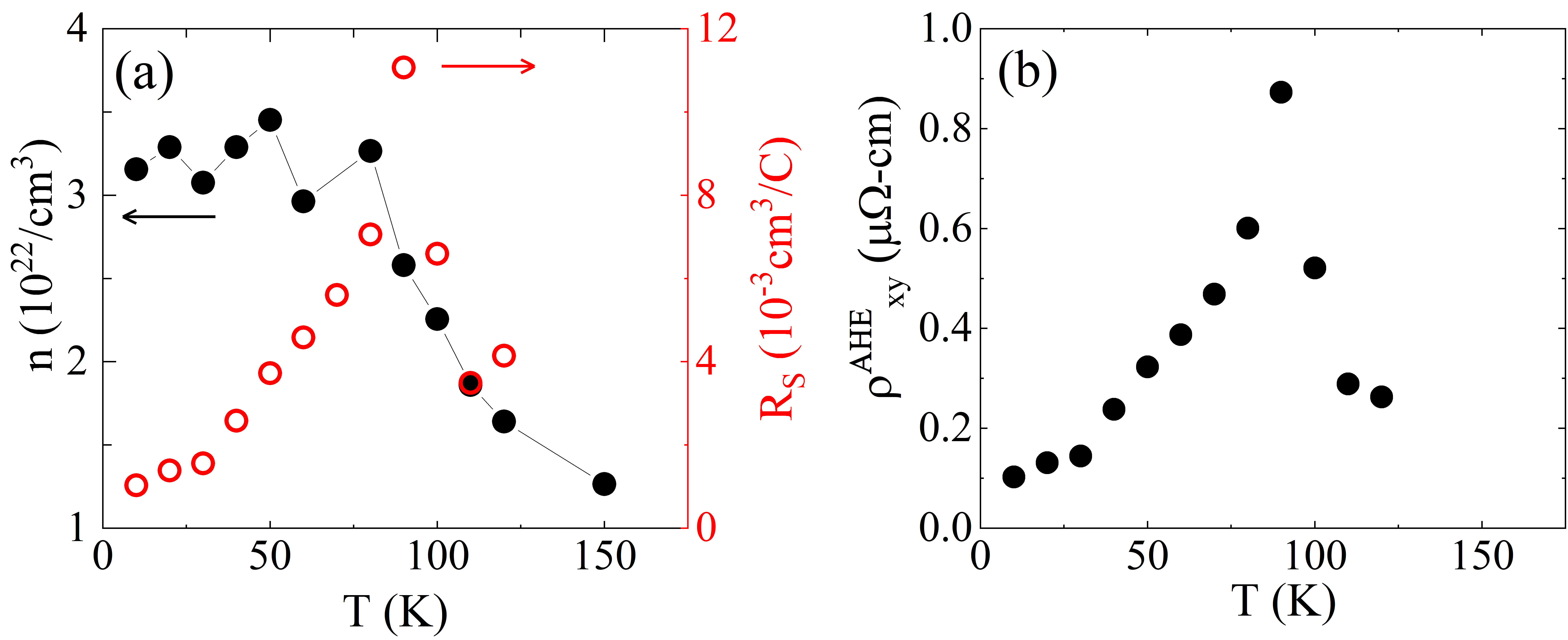}\hfill
\caption {(a) Temperature variation of the carrier concentration (left axis) and anomalous Hall coefficient R$_s$ (right axis). (b) Temperature variation of the anomalous Hall resistivity $\rho^{AHE}_{xy}$.}
\label{fig7}
\end{figure}

While the ordinary Hall effect, driven by the Lorentz force, is well understood, the AHE remains a topic of significant interest due to its complex origins and potential applications. The AHE can arise from both extrinsic mechanisms, like skew scattering and side jump effects influenced by spin-orbit interaction, and intrinsic mechanisms, such as the Karplus–Luttinger effect related to the spin-orbit interaction of Bloch electronic bands. Here, the measurements of Hall resistivity ($\rho_{xy}$) have been carried out over a wide field and temperature ($H-T$) phase space. Figure~\ref{fig6}(a) shows the $\rho_{xy} (H)$ data collected at several temperatures in the range of 10 to 140 K. A nonlinear magnetic field dependence of $\rho_{xy}$ is evident at T $\leq$ 120~K, which tends to reach saturation at a field above $\approx$ ± 0.8 tesla. The nonlinearity has its origin in the AHE which occurs alongside the conventional Hall effect. The transverse resistivity $\rho_{xy}$ of the sample measured in the standard Hall geometry is expressed as
\begin{equation}
\rho_{xy}= \rho^{OHE}_{xy}+\rho^{AHE}_{xy}
\label{eq3}
\end{equation}				 
The first term in this eq.~\ref{eq3} represents the ordinary Hall resistivity, caused by the Lorentz force on charge carriers and depends on the field as $R_0$$\mu_{0}$$H$, where $R_0$ is the Hall coefficient. The $R_{\rm 0}$ is negative over the entire temperature region, indicating that the dominant carriers are electron-type. The electron carrier density n can be deduced using the relation of n $\sim$ -1/(e$R_0$) which is shown in Fig.~\ref{fig7}(a), and it reaches a value 3.17$\times$10$^{22}$ cm$^{-3}$ at 10~K, which is of the same order as for other half-metallic systems, such as Co$_3$Sn$_2$S$_2$~\cite{wang2018large}. The carrier density remains almost temperature independent up to $\approx$ 90~K and beyond that it decreases with the increasing temperature. The anomalous Hall resistivity $\rho^{AHE}_{xy}$ in collinear ferromagnets has its origin in the magnetization $M$ of the sample and expressed as 4$\pi$$R_s$$M_{\rm s}$, where the $R_{\rm s}$(T) is anomalous Hall coefficient. The $R_{\rm s}$ increases monotonically with increasing temperature, reaching the maximum value at 90~K, and beyond this, it decreases as seen in Fig.~\ref{fig7}(a). 

The manipulation of the AHE by controlling magnetization is of great interest in condensed matter physics due to its potential applications in the design of spintronic devices. To verify that the pronounced AHE observed in this system is due to its ferromagnetic spin structure, we conducted magnetization measurements with the magnetic field aligned parallel to the film normal.

The $M$ vs field plots of the sample measured at the same set of temperatures where the $\rho_{xy}(H)$ was measured are shown in Fig.~\ref{fig6}(b). As for the $\rho_{xy}$, the magnetization also tends towards a constant value at $\approx$ ± 0.8 tesla. Furthe, the $M-H$ plots reveal a tinny loop with a coercive field of 150 Oe indicating the soft magnetic behavior of CoS$_2$, as shown in Fig.~\ref{fig6}(b). 

\begin{figure}[ht!]
\centering
\includegraphics[width=0.49\textwidth]{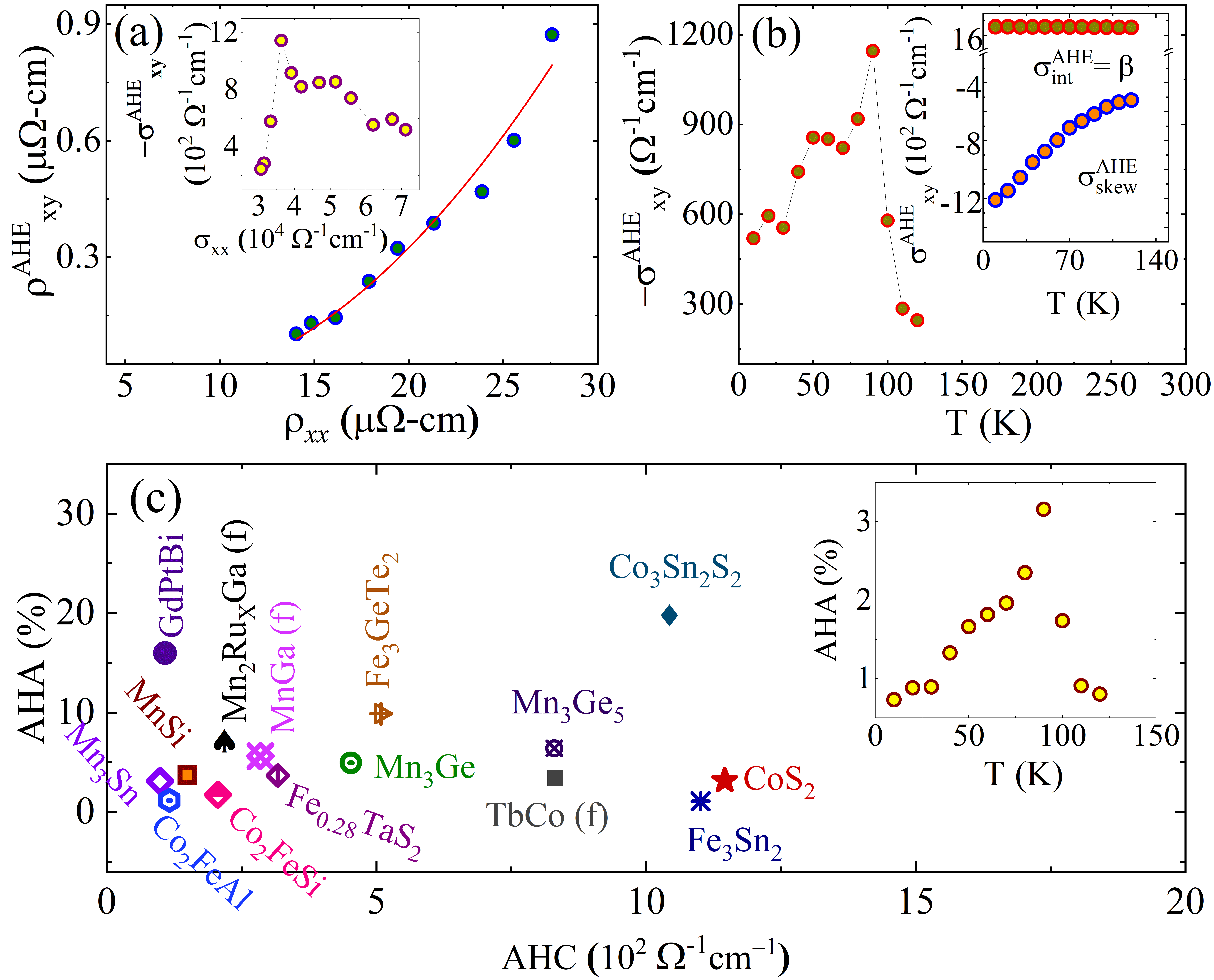}\hfill
\caption {(a) Fitting of anomalous Hall resistivity ($\rho^{AHE}_{xy}$) vs longitudinal resistivity ($\rho_{xx}$) using conventional scaling relation (Red line) and the variations of anomalous Hall conductivity (AHC) vs longitudinal conductivity (inset).  (b) The AHC as a function of temperature, and the variations of different AHC as a function of temperature (inset). (c) Comparison of anomalous Hall angle (AHA) and AHC values of CoS$_2$ with the reported results of other crystals exhibiting AHE. The temperature dependence of AHA (inset).}
\label{fig8}
\end{figure}

The field-dependent $\rho^{AHE}_{xy}$ at different temperatures has been calculated by subtracting the ordinary Hall contribution from the field-dependent $\rho_{xy}$ following Eq.~\ref{eq5}. Figure~\ref{fig7}(b) shows the temperature dependence of $\rho^{AHE}_{xy}$, which increases with temperature, reaching a maximum at 90 K. To quantitatively assess the contributions of intrinsic and extrinsic factors to the anomalous Hall resistivity, we fitted the $\rho^{AHE}_{xy}$ versus $\rho_{xx}$  data (blue spheres in Fig.~\ref{fig8}(a)) using the conventional scaling relation~\cite{tian2009proper, wang2018large}:
\begin{equation}
\rho^{AHE}_{xy}=\alpha\rho_{xx}+\beta\rho^{2}_{xx}				
\label{eq4}
\end{equation}		 
where $\alpha$ and $\beta$ provide information about extrinsic skew scattering and the combined effect of side jump scattering and intrinsic contributions arising from the band structure~\cite{karplus1954hall}, respectively. From this fitting, we obtained $\alpha$ $\approx$ -0.017 and $\beta$ $\approx$ 1660 $\Omega^{-1}$cm$^{-1}$ . The large value of $\beta$ indicates that the AHE is predominantly driven by intrinsic Karplus – Luttinger contribution ~\cite{karplus1954hall}, which is often described in terms of Berry Phase physics, and the side jump mechanism. 

The AHC $\sigma^{AHE}_{xy}$ of a collinear ferromagnetic scales with the longitudinal conductivity ($\sigma_{xx}$)~\cite{nagaosa2010anomalous}. The inset of Fig.~\ref{fig8}(a) presents AHC as a function of $\sigma_{xx}$, revealing a significant deviation from linearity. Such pronounced deviations have also been observed in magnetic Weyl metals~\cite{nakatsuji2015large}. We calculated the temperature-dependent $\sigma^{AHE}_{xy}$ using the following relation,
\begin{equation}
\sigma^{AHE}_{xy} =\frac{-\rho^{AHE}_{xy}}{{\rho^{AHE}}^{2}_{xy} + \rho^{2}_{xx}}	
\label{eq5}
\end{equation}	
which is shown in Fig.~\ref{fig8}(b). The AHC increases to 1145 $\Omega^{-1}$cm$^{-1}$ at approximately 90 K and then drops. The $\sigma^{AHE}_{xy}$ is a function of $\alpha$ and $\beta$ through Eq.~\ref{eq4} and \ref{eq5}. The side jump contribution to $\sigma^{AHE}_{xy}$ can be estimated using the expression ($e^2$/($ha$)(E$_{\rm so}$/E$_{\rm F}$), where E$_{\rm so}$ and E$_{\rm F}$ are the spin-orbit interaction energy  and the Fermi energy respectively ~\cite{onoda2006intrinsic, nozieres1973simple}. The physical quantities $e$, $h$, and $a$ are the electronic charge, Planck’s constant, and lattice parameter, respectively. For most ferromagnetic metals, E$_{\rm so}$/E$_{\rm F}$ is of the order of 10$^{-2}$~~\cite{wang2018large}, Using the lattice constant $a$ $\approx$~5.59~\AA, the derived extrinsic side-jump contribution is only about 6.92 $\Omega^{-1}$cm$^{-1}$, which is much smaller compared to the intrinsic part. Using the values of $\alpha\rho_{xx}$ and $\beta$, we calculated the temperature-dependent magnitudes of extrinsic skew scattering AHC ($\left|\sigma^{AHE}_{skew}\right| = (\alpha\rho_{xx}$)/$\rho_{xx}^2$ = $\alpha\ /\rho_{xx}$) and of the intrinsic AHC ($\left|\sigma^{AHE}_{int}\right| = \beta$). These results are plotted in the inset of Fig.~\ref{fig8}(b). Here, we also calculate anomalous Hall angle (AHA = $\sigma^{AHE}_{xy}/\sigma_{xx}$), which is shown as inset of Fig.~\ref{fig8}(c) The AHA is an intrinsic parameter, which defines the efficiency of the conversion of longitudinal current into transverse current or, anomalous Hall current. The maximum value of AHA (\%) is seen at $\approx$ 90 K. The results of AHC are shown in Fig.~\ref{fig8}(c) and compared with reported values for several non-collinear magnets. Here, the result for the current compound are represented by the red star symbol in Fig.~\ref{fig8}(c). Our sample exhibits the highest AHC and a significant AHA (\%) among all noncollinear magnets. The high AHC in this material is a consequence of a large Berry phase (discussed below). 

\begin{figure}[ht!]
\centering
\includegraphics[width=0.45\textwidth]{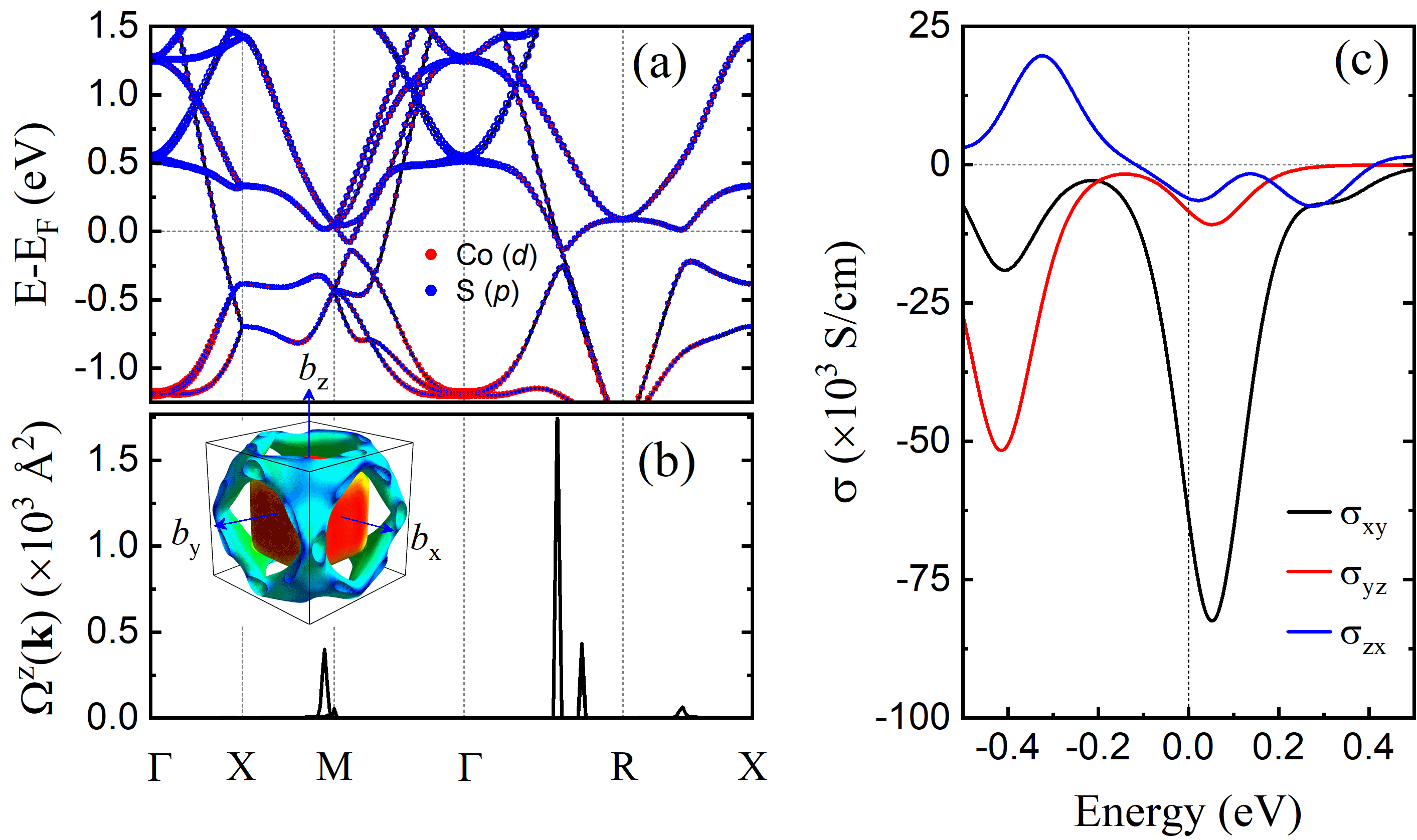}
\caption {(a) Atomic projected electronic band structure, (b) Berry curvature $\Omega^z(k)$ (in $10^3$ \AA$^2$) along the high-symmetry path, with an inset depicting the three-dimensional Fermi surface, and (c) anomalous Hall conductivity components $\sigma_{xy}$, $\sigma_{yz}$, and $\sigma_{zx}$ (in $10^3$ S/cm) near the E$_{\rm F}$ as a function of energy.}
\label{dft}
\end{figure}

\HP {To understand the electronic structure origin of the AHE, we calculate the band structure, Fermi surface and Berry curvature. In Fig.~\ref{dft}(a), we  show calculated atom projected electronic band structure of CoS$_2$. These hybrid functional band structure calculations include SOC. The bands near the E$_{\rm F}$ primarily consist of the $p$-states of sulfer, strongly hybridized with Co $d$-states. Inclusion of SOC leads to splitting of bands particularly along the $\Gamma$-R direction due the unpaired electrons in Co $d$ orbitals. A pronounced peak in Berry curvature, $\Omega^z(\textbf{k})$, is observed in Fig.~\ref{dft}(b), which results from the nearly degenerate band splitting induced by SOC along the $X$–$M$ and $\Gamma$–$R$ directions. Notably, the uniform sign of $\Omega^z(\textbf{k})$ throughout the Brillouin zone suggests a net Berry curvature flux and points toward the presence of topologically non-trivial band structures, potentially supporting robust edge states. This intrinsic Berry curvature acts as a fictitious magnetic field in momentum space, thereby contributing directly to the AHC. As a consequence, the calculated AHC, $\sigma_{zx}$, shown in Fig.~\ref{dft}(c), exhibits a strong energy dependence in the vicinity of the E$_{\rm F}$. The sign reversal in $\sigma_{zx}$ as the E$_{\rm F}$ is tuned reflects the competing contributions of Berry curvature from different regions of the Brillouin zone. This behavior underscores the sensitivity of the AHE to subtle details in the band topology and highlights the potential for tuning the Hall response via doping or gating.}

\section{Conclusions}
\HP{In conclusion, the experimental measurements of electronic transport and magnetization combined with \textit{ab initio} calculations confirm the half-metallic nature of CoS$_2$. Electronic structure calculations also show that application  of a small strain transforms this half metallic characteristic to metallic. A pronounced cusp is seen in the field dependence of the magnetoresistance at $T$ $\leq$ 80~K, which appears to be a result of the competition between field suppression of weak localization and carrier back scattering by randomly oriented magnetic domains. A transition from negative to positive magnetoresistance is also seen above the ordering temperature. Additionally, the experimental observation of anomalous Hall resistivity, along with \textit{ab initio} computed band structure and Berry curvature, suggest that the AHE is primarily driven by the intrinsic Karplus–Luttinger contribution, which is often associated with Berry phase physics.}

\section*{Acknowledgment}
This research has been performed at the United States Department of
Defense funded Center of Excellence for Advanced Electro-photonics
with 2D materials – Morgan State University, under grant No. W911NF2120213. The theoretical and computational research is supported as part of the Center for Energy Efficient Magnonics, an Energy Frontier Research Center funded by the U.S. Department of Energy, Office of Science, Basic Energy Sciences under Award number DE-AC02-76SF00515. H.P and D.P acknowledge the use of the computational facilities on the Frontera supercomputer at the Texas Advanced Computing Center (TACC) via the pathway allocation, DMR23051. We also thank to Prof. Rajeswari Kolagani and Marcus Rose from Towson University for their assistance with the X-ray diffraction measurements.

\bibliography{references}
\end{document}